\documentclass[final, 5p, times, twocolumn, 10pt]{elsarticle}
\usepackage{graphicx}
\usepackage{latexsym}
\usepackage{amsmath,amssymb,algorithmic,algorithm}
\usepackage{slashbox}

\usepackage[tight,footnotesize]{subfigure}
\newcommand{\para}[1]{{\vspace{5pt} \noindent \bf #1 \hspace{4pt}}}

\title{Cross-Band Interference Considered Harmful in OFDM Based Distributed Spectrum Sharing}

\author[label1]{Wei Hou\fnref{label3}}
\author[label1]{Lin Zhang}
\author[label2]{Lei Yang}
\author[label2]{Heather Zheng}
\author[label1]{Xiuming Shan}

\address[label1]{Department of Electronic Engineering, Tsinghua University, Beijing 100084, P. R. China}
\address[label2]{Department of Computer Science, University of California at Santa Barbara, Santa Barbara, CA 93106, USA} 

\fntext[label3]{Corresponding author. Tel.:+86-10-62797587.\\
E-mail address: hou-w05@mails.tsinghua.edu.cn (Wei Hou).}

\begin{document}

\begin{frontmatter}

\begin{abstract}
In the past few years we have witnessed the paradigm shift from static
spectrum allocation to dynamic spectrum access/sharing. 
Orthogonal Frequency-Division Multiple Access (OFDMA) is a promising
mechanism to implement the agile spectrum access. However, in wireless distributed
networks where tight  synchronization is infeasible, OFDMA faces the
problem of cross-band interference. Subcarriers used by different users 
are no longer orthogonal, and transmissions operating on non-overlapping subcarriers can
 interfere with each other. In this paper, we explore the cause of
cross-band interference and analytically quantify its strength and impact on
packet transmissions. Our analysis captures three key practical
artifacts:  inter-link frequency offset, temporal sampling mismatch
 and power heterogeneity. 
   To our best knowledge, this work is the
first to systematically analyze the cause and impact of cross-band interference. 
Using insights from our analysis, we then build and compared three mitigating methods to
combat cross-band interference.
Analytical and simulation results show that placing  frequency guardband  at
link boundaries is the most effective solution in distributed spectrum sharing, 
while the other two frequency-domain 
methods are sensitive to either temporal sampling mismatch 
or inter-link frequency offset. We find that the proper guardband size depends heavily on
power heterogeneity. Consequently, protocol
designs for dynamic spectrum access should carefully take into account the cross-band interference
when configuring spectrum usage. 
\end{abstract}


\end{frontmatter}

\section{Introduction}
\label{sec:intro}
In the past few years we have witnessed the paradigm shift from static
spectrum allocation to dynamic spectrum access/sharing~\cite{cr_survey,cr_survey2}.  
In this new model, wireless
devices no longer operate on statically assigned spectrum, but
acquire spectrum on-demand and share with peers in their local
neighborhood. Now wireless devices can obtain spectrum they really
need,  while improving spectrum utilization through
multiplexing.

To realize dynamic spectrum access, one must enable wireless radios to configure and
adapt their frequency usage on-the-fly.  The widely-known radio solution is  orthogonal
frequency division multiple access~(OFDMA)~\cite{OFDMA,DSA_Wimax,DSA_OFDMA07,Yang08-a}, 
where a large spectrum band is divided into many
frequency subcarriers and an OFDMA radio can access spectrum dynamically by
communicating on any combination of the subcarriers.
Spectrum multiplexing is achieved by assigning non-overlapping sets of the subcarriers
to different users. Using the IDFT/DFT (inverse discrete Fourier transform/discrete Fourier transform) based modulation, all the subcarriers appear
orthogonal to each other. Due to its good flexibility and simplicity in implementation,
OFDMA has already been applied in IEEE 802.16e (WIMAX)~\cite{802.16} standard. 

OFDMA radios, however, are highly sensitive to errors in time and frequency
synchronization. Imperfect synchronization destroys
the orthogonality among subcarriers and results in inter-carrier interference (ICI)~\cite{OFDMA-syncerr}.
In centralized networks like WiMAX, an effective way for mitigating ICI is to improve the time and frequency synchronization 
precision~\cite{OFDMA_EST1,OFDMA_EST2,JointSync05}.
And in MIMO-OFDM system, in addition to ICI, cross-antenna interference~\cite{Weizhang07,Tang08} 
is also one of the main concerns.
While in distributed networks, the concurrent asynchronous transmissions
produce harmful interference to each other. Moreover,
the out-of-band emissions (OOB) and the asynchronism induced {\em cross-band interference} 
cannot be simply filtered out as it leaks into the useful signal's spectrum~\cite{Tang08,hou09}. 
The problem of OOB mitigation in OFDM systems
has been studied, and time/frequency domain  mitigating methods have been proposed.
Windowing based  methods~\cite{Window,AST} mitigate the signal's power leakage in time-domain,
while frequency domain methods deal with this problem
by either inserting frequency guardband~\cite{Window} or 
performing cancellation coding~\cite{CC,PCC,WMMSE} or 
using multiple choice sequences approach (MCS)~\cite{MCS}.
However, most existing works analyze the unwanted power leakage  
from the view of the transmitter or its corresponding receiver,
which may be different from the view of other receivers as links are asynchronous.

In this paper, we consider the distributed spectrum sharing in
wireless systems where no central control is present and  tight time
synchronization becomes infeasible.  
In this case, we show that cross-band
interference can be highly harmful and lead to large performance
degradation. 
  To better understand this special type of interference, 
we develop a systematic framework to evaluate its impact 
on asynchronous OFDMA transmissions from the receiver's point of view. 
We seek to 
understand in detail the origin and characteristics
of  the cross-band interference, examine its impact on packet detection and 
transmission, and identify effective solutions to suppress the interference.
This work builds on our prior work~\cite{hou09} that preliminarily examines the strength of
cross-band interference, but makes more comprehensive analysis on
the cross-band interference and extends to examine how cross-band interference
affects preamble detection and packet reception in the presence of channel
fading.

We use a two-step process to analyze cross-band interference. We start from
deriving the amount of interference a subcarrier can produce on neighboring
subcarriers occupied by other transmissions. Three key 
practical artifacts are captured: inter-link frequency offset, temporal sampling mismatch
 and  power heterogeneity.  
Our analysis produces the statistically average
interference strength experienced by any frequency subcarrier. 
Next,  we  analyze
the impact of cross-band interference on packet level performance by
examining how cross-band interference affects packet
synchronization.

Built on the theoretical analysis, we further propose the cross-symbol
cancellation (CSC), a new frequency-domain interference cancellation approach,
to address the interference in the presence of large temporal mismatch.
We then examine and compare three mechanisms for tackling cross-band interference:
frequency guardband (FGB), inter-symbol cancellation (ISC) and CSC.
By adding frequency redundancy or overhead to each transmission, 
these solutions seek to reduce the
interference at its source or insert robustness at each receiver
against the interference.
We examine how these mechanisms perform in distributed networks
in the presence of temporal mismatch and inter-link frequency offset.
We then evaluate their effectiveness and confirm our conclusions using simulations.

Our work makes two key contributions:

\begin{itemize}
\item We build an analytical model to characterize the impact of cross-band
  interference on OFDMA transmissions. 
Our work is the first to analytically evaluate
its impact on asynchronous OFDMA transmissions, taking into account the
packet synchronization and data reception. Our analytical results
closely match the experimental results.
\item We propose a new  approach--CSC to tackle cross-band interference. 
Using both theoretical analysis and
simulation experiments, we compare three mitigating methods and show that ISC is ineffective due to
temporal mismatch, and CSC is sensitive to frequency offset. Overall, inserting
FGB between transmissions is the most efficient solution in distributed 
spectrum access.
\end{itemize}

The rest of the paper is organized as follows. In Section~\ref{sec:model}
we briefly introduce the preliminaries of OFDMA and asynchronous OFDMA.
We then in Section~\ref{sec:oob}  characterize the strength of
cross-band interference and in
Section~\ref{sec:impact} study its impact 
on OFDMA operations especially packet
synchronization. We propose and
analyze three approaches to tackle the interference in
Section~\ref{sec:csc}. 
Simulation results are presented in Section~\ref{sec:eval} and conclusions are
drawn in Section~\ref{sec:conclusion} finally.

\section{Preliminaries}
\label{sec:model}
As background, we  briefly discuss the basic operation of OFDMA and the problems it faces when
transmissions are asynchronous.

\subsection{OFDMA}
In OFDMA, each transmitter uses OFDM to transmit signals
over their selected subcarriers. The transmitter maps the
modulated bit stream into the selected subcarriers, 
and applies IDFT to convert 
the bit stream into time-domain OFDM symbols. In this way,
it only ``pours'' power over the selected subcarriers.  
To decode the OFDM signal, 
the receiver extracts and adjusts the time-domain
symbols through synchronization, and applies DFT to
reconstruct the bit stream from the intended subcarriers.

Mathematically, the IDFT and DFT operations can be expressed as: 
\begin{equation}
\left\{
	\begin{aligned}
    &t(n)= IDFT\big[s(k)\big] = \sum_{k\in \Omega}s(k)e^{i2\pi \frac{kn}{N}}\\
	&s(k) = DFT\big[t(n)\big] = \sum_{n=0}^{N-1}t(n)e^{-i2\pi \frac{kn}{N}}
	    \end{aligned}
	\right.
\end{equation}
where $N$ is the number of IDFT/DFT points which is often set as the power of $2$ 
so that fast Fourier transform (FFT) and inverse FFT (IFFT) can be used to achieve efficiency.
$s(k)$ denotes the frequency-domain data on the $k$th subcarrier, $t(n)$ is
the $n$th sampling point of time-domain OFDM symbol, and
$\Omega$ is the set of occupied subcarriers. 
Different links possess non-overlapping $\Omega$s, so multiple transmissions
can take place simultaneously in time and be distinguished in frequency.

One important requirement in OFDMA is to maintain the subcarrier orthogonality
so that simultaneous transmissions from  different subcarriers
will not interfere with each other.  
When transmissions from coexisting links are
perfectly synchronized in time and frequency, the DFT operations
at their receivers will remove unwanted signals and maintain subcarrier orthogonality.

\begin{figure}[!t]
	\centering	
	\resizebox{0.48\textwidth}{!}{\includegraphics{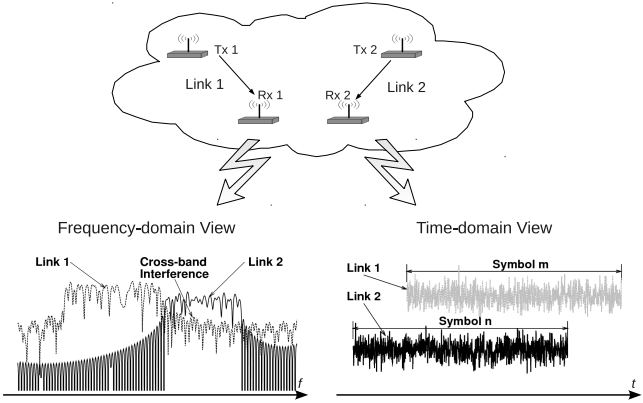}}	
	\caption{Illustrating the spectrum sharing in wireless distributed network.
	 The two OFDM transmissions are viewed at Rx 2 in both frequency domain and time domain. }
	\label{fig:ofdma}
\end{figure}

\subsection{Distributed OFDMA}
In distributed OFDMA, due to the inherent asynchronism, subcarrier orthogonality across 
 links is destroyed and harmful cross-band interference is created.
Figure~\ref{fig:ofdma} shows 
a distributed OFDMA network with two independent links--link 1 and link 2, each occupying 
a non-overlapping set of subcarriers. The frequency-domain and time-domain snapshots of the 
two transmissions viewed at link 2's receiver are shown on the bottom. 
In this paper, we characterize the distributed OFDMA and the cross-band interference 
by taking into account the following three artifacts:

\begin{itemize}
\item {\em Inter-link Frequency offset $(\epsilon)$} is the  
  central frequency discrepancy between the two links'
  transmitters.
  It keeps stable during a short period if the environmental parameters 
  (such as temperature, humidity) do not change rapidly. 
  We assume $|\epsilon|\le 0.5$ subcarrier in this paper. 
\item {\em Temporal mismatch $(\tau)$} refers to the difference between the two links'
  symbol arrival time viewed at the receiver, 
  as shown in Figure~\ref{fig:ofdma} (right bottom).
  We assume that $\tau$ is evenly distributed in $[0,T+T_{CP})$,  where 
  $T$ and $T_{CP}$ denote the symbol length (excluding the cyclic prefix) and the cyclic prefix length, respectively.
\item {\em Power Heterogeneity $(p_r)$} refers to the difference in average power per subcarrier
  of different transmissions observed at the receiver. The problem of power heterogeneity can be 
  alleviated using power control in cellular network~\cite{PowControl_JSAC03}, but remains 
  in distributed network as users are not coordinated.
  $p_r$ is defined as the power ratio of link 1 to link 2.
\end{itemize}

\section{Characterizing Cross-Band Interference}
\label{sec:oob}
In this section, we examine the cross-band interference in asynchronous
OFDMA systems. Using the example in Figure~\ref{fig:ofdma}, we treat link 1
as the interferer and examine its interference at link 2's
receiver in the presence of the temporal mismatch and inter-link
frequency offset. We use the rectangular pulse shaping for OFDM symbols,
and assume multipath fading channels of $K$-factor Rician distribution for both links,
where the $K$-factor is defined as the ratio of signal power in 
line-of-sight component over the scattered power.
Our study of two-link network can easily apply to the scenario of multiple links.

\begin{figure}[!t]
	\centering
    \resizebox{0.48\textwidth}{!}{\includegraphics{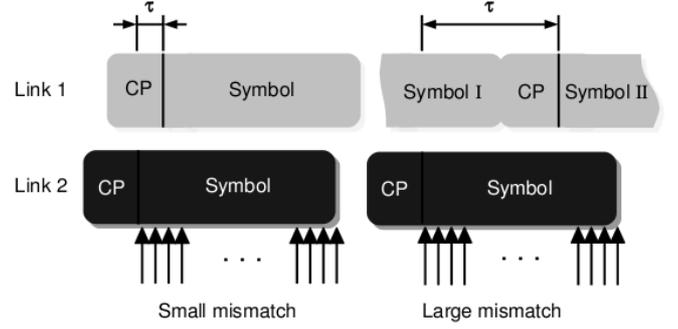}}
    \caption{Two temporal mismatch cases. Small 
    temporal mismatch happens when $\tau \le T_{CP}$ (left), while  large
	temporal mismatch happens when $  T_{CP} < \tau < T+T_{CP}$ (right). }
    \label{fig:sampling}
\end{figure}

As shown in
Figure~\ref{fig:sampling},  we divide our
analysis into two cases based on the degree of temporal mismatch: small mismatch when $ 0\le \tau
\le T_{CP}$ or large mismatch when $T_{CP}<\tau<T+T_{CP}$.
To simplify our analysis, we assume the users in this network are homogeneous with the
same OFDM parameters. We will examine the impact of user heterogeneity in our future work.

\subsection{Case A: Small Temporal Mismatch}
\label{sec:oob_a}
From Figure~\ref{fig:sampling}, when $0\le \tau \le T_{CP}$, each desired OFDM symbol of link 2 will face interference from only one
symbol of its interferer.  Let
$s_1(k)$ represent the frequency-domain data sent by link 1 on subcarrier
$k$. At link 2's receiver, it will experience a phase rotation in
the frequency domain due to the temporal mismatch $\tau$:
\begin{equation}
	\big(s_1(k)\big)_\tau \triangleq s_1(k)e^{-i 2\pi  \frac{k}{T}\tau }
	\label{eq:phase_shift}
\end{equation}
where $N$ is the total number of subcarriers.

To determine the cross-band interference from link 1 to link 2, we compute
the power spectrum of $\big(s_1(k)\big)_\tau$ together with the channel response
at any continuous frequency $f \in [0,N)$. This allows us
to determine the interference in the presence of inter-link frequency offset $\epsilon$,
 where link 1's transmissions at
subcarrier \#$i$ arrive at link 2's subcarrier \#$i+\epsilon$.
For example, $f=2$
refers to the frequency location of subcarrier \#$2$, and $f=2.5$ refers to the middle of subcarriers \#$2$ and \#$3$.

Let $\Omega_1$ represent the set of subcarriers occupied by the interferer
link 1.
 We perform discrete-time Fourier transform (DTFT) to derive
 ($P^A_{1\rightarrow 2}(f)$), link 1's
power spectrum seen by link 2  at any continuous $f$:

\begin{equation}
P^A_{1\rightarrow 2}(f)=|H_{1\rightarrow 2}(f)|^2 \cdot  |S_1(f)|^2
\label{eq:eq_A0}
\end{equation}
where $H_{1\rightarrow  2} (f)$ is the
frequency-domain channel response  between link 1 and 2,
$S_1(f)$ is
the interferer's spectrum represented as:

\begin{equation}
S_1(f)\!\!=\!\!\sum_{k\in\Omega_1}\!\!\big(s_1(k)\big)_\tau\!\frac{\sin[\pi(f\!\!-\! k)]}{N\sin[\frac{1}{N}\pi(f\!\!-\!k)]}
 e^{-i\pi(f\!-\!k)\frac{N\!-\!1}{N}}.
\label{eq:eq_A}
\end{equation}

\begin{figure*}[!t]
	\centering
	\subfigure[Time-frequency domain view] {\resizebox{0.48\textwidth}{!}{\includegraphics{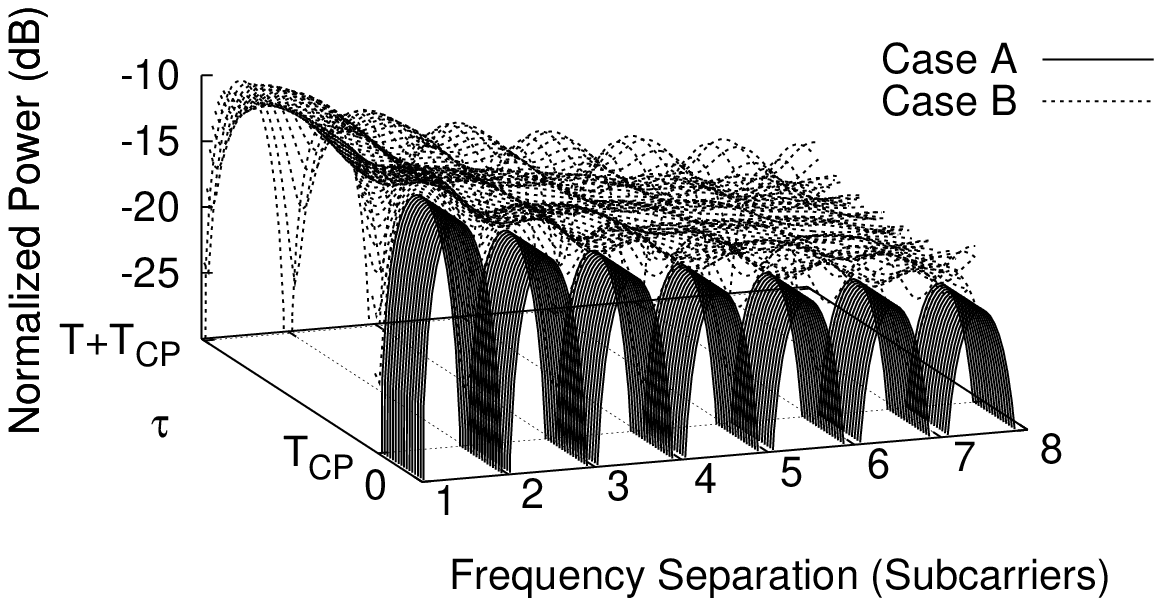}}}
	\hfill
	\subfigure[Frequency domain view]{\resizebox{0.48\textwidth}{!}{\includegraphics{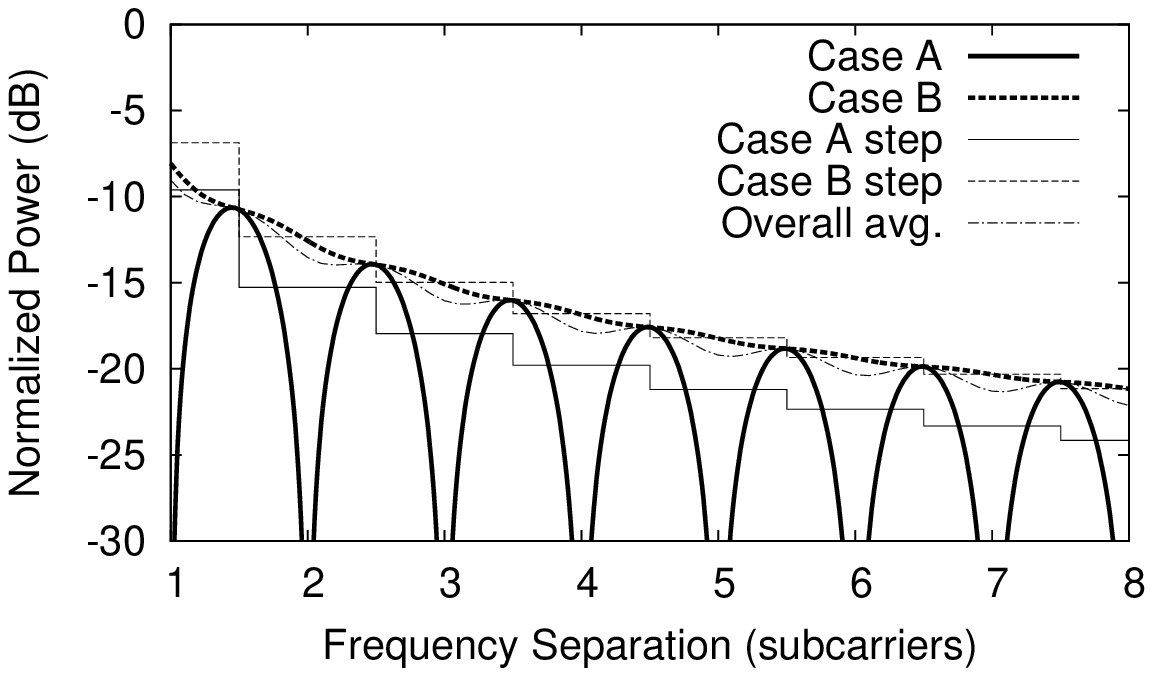}}}
    \caption{Analytical average power of the cross-band interference (normalized by $P_1$). (a) Viewed in time-frequency domain;
    (b) Viewed in frequency domain after averaging over the temporal mismatch.
    Assuming $T_{CP}=T/4$, and link 1 operates on 8 subcarriers.}
\label{fig:comp_sidelobe}
\end{figure*}

The derivation of Equation~(\ref{eq:eq_A}) can be found in  \textit{Appendix A}.
It is easy to show
that  $P_{1\rightarrow 2}^A(f) =0$ at integer
$f$ {($f\notin\Omega_1$) (see Figure~\ref{fig:comp_sidelobe}(a)}).
This means that the cross-band interference can be
fully canceled if $\epsilon$ is zero.

\subsection{Case B: Large Temporal Mismatch}

When $T_{CP}<\tau<T+T_{CP}$, each desired OFDM symbol of link 2 will face interference
from two truncated symbols of link 1 (see
Figure~\ref{fig:sampling}). Without loss of generality, we assume it
overlaps with
symbol~I in $M$ sampling points and symbol~II in $N-M$ 
points, where $M$ is determined by 
the temporal mismatch ($M=\lceil(\tau-T_{CP})\cdot N/T\rceil$, 
where $\lceil \cdot \rceil$ refers to rounding up the argument to the nearest integer). Let
$s_{1}^{I}(k)$ and $s_{1}^{I\!I}(k)$ represent the
frequency-domain constellation points on the $k$th subcarrier 
from Symbol I and Symbol II respectively. 
Due to the
temporal mismatch, each signal goes through 
a phase rotation and a truncation based on the non-mismatch symbol where $\tau=0$:
first, each symbol shifts its sampling location by $\tau$ to the right,
which corresponds to $s_{1}^{I}(k)$'s phase rotation by $-2\pi k\tau$, and then
Symbol I removes the tail $N-M$ points and 
Symbol II removes the front $M$ points.

The interfering signal's power spectrum becomes:
\begin{equation}
	P^B_{1\rightarrow 2} (f) =|H_{1\rightarrow 2}(f)|^2 \cdot |\big(S_{1}^{I}(f)+S_{1}^{I\!I}(f)\big)|^2
	\label{eq:eq_B0}
\end{equation}
where $S_{1}^{I}(f)$ and $S_{1}^{I\!I}(f)$ are the DTFT outputs of the two OFDM symbols respectively:

\begin{equation}
	\label{eq:eq_B}
\left\{
 \begin{aligned}
  S_{1}^{I}(f) & = \sum_{k\in \Omega_1} \big(s_{1}^{I}(k)\big)_{\tau}  \frac{\sin[\frac{M}{N}\pi(f-k)]}{N\sin[\frac{1}{N}\pi(f-k)]}
					\\ &\qquad\cdot e^{-i\pi(f-k)\frac{M-1}{N}}\\
  S_{1}^{I\!I}(f) & = \sum_{k\in\Omega_1}\big(s_1^{I\!I}(k)\big)_{\tau}  \frac{\sin[\frac{N-M}{N}\pi(f-k)]}{N\sin[\frac{1}{N}\pi(f-k)]}
					\\ &\qquad\cdot e^{-i\pi(f-k)\frac{N+M-1}{N}}.
 \end{aligned} \right.
\end{equation}
The derivation of Equation~(\ref{eq:eq_B}) can be found in 
 \textit{Appendix A}. 
Because $s_{1}^{I}(k)$ and $s_{1}^{I\!I}(k)$ are 
randomly chosen from the constellation map and independent of each other, the power spectrum 
becomes randomly distributed across $f$ and
is no longer zero at integer $f$ locations.

From the above analysis, we see that the cross-band interference
at any specific $f$
depends heavily on the channel property $H_{1 \to 2}(f)$, temporal mismatch level $\tau$ and the configuration of
signal constellation $s(\cdot)$. 
In the following, we perform statistical analysis to derive the average interference
strength.

\subsection{Statistical Analysis}
To characterize the cross-band interference, we make a statistical analysis 
to achieve the average cross-band interference strength.
Our analysis makes the following assumptions. 
$s_1(k)$, $s_{1}^{I}(k)$ and $s_{1}^{I\!I}(k)$ are independent and identically distributed (i.i.d)
with constant average power $E[|s_1(k)|^2]=P_1$ for all $k \in \Omega_1$,
and the average channel gain  is unified, \textit{i.e.}, $E\big[|H_{1\to2}(f)|^2\big]=1$.

\para{Time-Frequency Domain View.}
We first derive the average interference strength at any
frequency location and temporal mismatch pair $(f,\tau)$ by averaging over
 $H_{1 \to 2}(f)$ and $s_1(k)$.

For Case A,  with respect to Equation~(\ref{eq:eq_A0}), 
by averaging over the statistical distribution of $H_{1 \to 2}(f)$ and $s_1(k)$,
we get the average cross-band interference strength:
\begin{equation}
\begin{aligned}
  P_{CBI}^A(f,\tau)&=E\big [P^A_{1\rightarrow 2}(f)\big] \\ &
  = P_1\sum_{k\in \Omega_1} \frac{\sin^2[\pi(f-k)]}{N^2\sin^2[\frac{1}{N}\pi(f-k)]} .
  \label{eq:eq_av_At}
  \end{aligned}
\end{equation}
We observe that the average cross-band interference under small temporal
mismatch ($\tau\in[0,T_{CP}]$) is independent  of $\tau$. 
This property has been utilized in
centralized OFDMA networks that apply cyclic prefix to reduce the impact of
imperfect timing synchronization.

For Case B, by averaging over  $s_1^I(k)$ and $s_1^{I\!I}(k)$ (which are
independent of each other) as well as $H_{1\to 2}(f)$ in Equation~(\ref{eq:eq_B0}), 
the  average interference strength becomes:

\begin{equation}
	\label{eq:eq_av_Bt}
	\begin{aligned}
    &P_{CBI}^B(f,\tau)=E\big [P^B_{1\rightarrow 2}(f)\big ] \\ 
	&\!\!=P_1 \sum_{k\in \Omega_1}\frac{\sin^2[\frac{M}{N}\pi(f-k)]+\sin^2[\frac{N-M}{N}\pi(f-k)]}{N^2\sin^2[\frac{1}{N}\pi(f-k)]} \\
	&\!\!\approx \!P_1 \!\!\!\sum_{k\in \Omega_1} \!\! {\frac{\sin^2[\frac{\tau\!-\!T_{CP}}{T}\pi(\!f\!\!-\!\!k)]\!\!+\!\sin^2[\frac{T\!-\!(\tau\!-\!T_{CP})}{T}\pi(\!f\!\!-\!\!k)]}{N^2\sin^2[\frac{1}{N}\pi(f\!-\!k)]}}.
	    \end{aligned}	  
\end{equation}

Figure~\ref{fig:comp_sidelobe}(a) shows the average interference 
strength (normalized by $P_1$) at any ($f$,$\tau$) pair. 
As expected, $\tau=T_{CP}$  marks an obvious boundary between Case A and B.

\para{Frequency Domain View.}
To examine the average cross-band interference over time,  
we take an average over the temporal mismatch $\tau$. 
Note that $\tau$ follows a random uniform distribution in $[0, T+T_{CP})$ as different transmissions
may start at random time.

For Case A, Equation~(\ref{eq:eq_av_At}) is already independent of $\tau$, thus we have

\begin{equation}
  P_{CBI}^A(f) = P_1\sum_{k\in \Omega_1} \frac{\sin^2[\pi(f-k)]}{N^2\sin^2[\frac{1}{N}\pi(f-k)]}.
  \label{eq:pcbi_A}
\end{equation}

For Case B, the average cross-band interference strength at any frequency location $f$
is the integral of $P^B_{CBI}(f,\tau)$ (in Equation~(\ref{eq:eq_av_Bt})) with respect to $\tau$:
\begin{equation}
\begin{aligned}
    P_{CBI}^B(f)&=\frac{1}{T}\int_{T_{CP}}^{T+T_{CP}} P^B_{CBI}(f,\tau) \mathrm{d}\tau \\ &
	\approx P_1 \sum_{k\in \Omega_1}\frac{1-\frac{\sin[2\pi(f-k)]}{2\pi(f-k)}}{N^2\sin^2[\frac{1}{N}\pi(f-k)]}.
	\label{eq:pcbi_B}
	    \end{aligned}
\end{equation}

Figure~\ref{fig:comp_sidelobe}(b) shows the average cross-band 
interference strength as a function of frequency separation.
To draw a comparison between the cases of small mismatch (Case A) and large mismatch (Case B),
we average the interference strength 
across one subcarrier's span, \textit{i.e.}, averaging over $f$ in $(0.5, 1.5)$, $(1.5, 2.5)$ and so on.
The resultant curves labelled by "step" in Figure~\ref{fig:comp_sidelobe}(b) show that 
the average cross-band interference strength of
Case B is stronger by approximately $3$ dB than that of Case A.

By combining the results of Case A and B statistically, 
we derive the overall average interference
strength:
\begin{equation}
\begin{aligned}
	&P_{CBI}(f)=\rho\cdot P_{CBI}^A(f) + (1-\rho) \cdot P_{CBI}^B(f)\\
	&\approx \!\!P_1\!\!\sum_{k\in \Omega_1}\!\! \frac{\frac{T_{CP}}{T+T_{CP}}\sin^2[\pi(f\!\!-\!\!k)]\!+\!\frac{T}{T+T_{CP}} [ 1\!\!-\!\!\frac{\sin[2\pi(f-k)]}{2\pi(f-k)}]}{N^2\sin^2[\frac{1}{N}\pi(f-k)]}
\end{aligned}
	\label{eq:pcbi}
\end{equation}
where $\rho$ ($\rho=\frac{T_{CP}}{T+T_{CP}}$) and $1-\rho$ are 
the event  probabilities of Case A and B.
Note that $\rho$ is equivalent to the additional overhead due to cyclic prefix. 
Using this result, we
are able to evaluate the expected cross-band interference strength at 
given frequency locations statistically. We also plot
the overall average cross-band interference strength in Figure~\ref{fig:comp_sidelobe}(b).

\begin{figure}[!t]
	\centering
	\subfigure[Impact of interferer's subcarriers \# ($L$.) 
		]{\resizebox{0.48\textwidth}{!}{\includegraphics{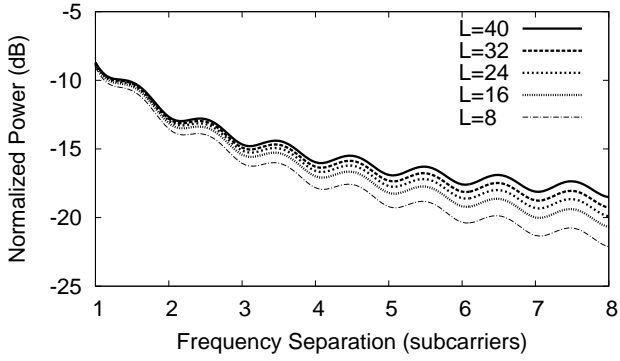}}}
	\vfill
	\subfigure[Impact of cyclic prefix's overhead ($\rho$).
		]{\resizebox{0.48\textwidth}{!}{\includegraphics{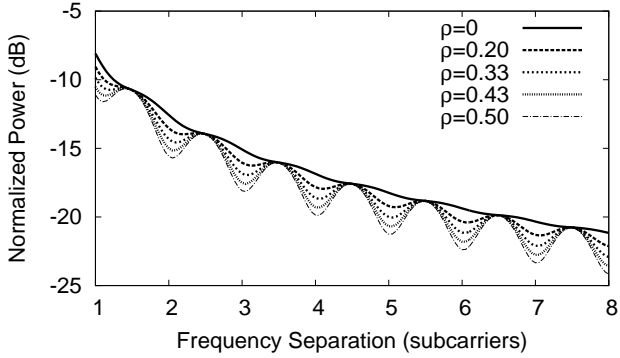}}}
	\vfill
	\subfigure[Impact of DFT/IDFT points \# ($N$).
		]{\resizebox{0.48\textwidth}{!}{\includegraphics{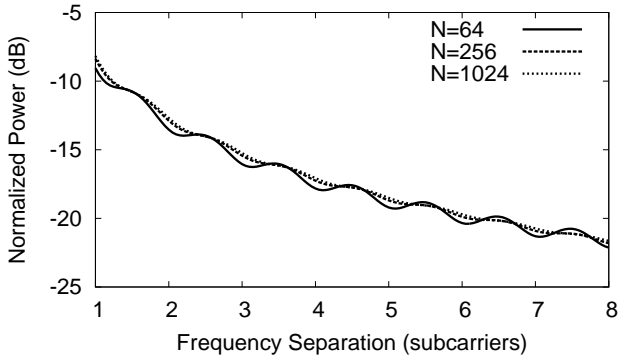}}}
    \caption{Examining the impacts of three parameters $(L,\rho,N)$ to the average cross-band interference strength. 
    	We set $L=8$, $\rho=0.2, N=64$ by default. (a) Impact of $L$; (b) Impact of $\rho$;
    	(c) Impact of $N$ (the two curves of $N=256$ and $N=1024$ are overlapping).}
\label{fig:comp_3para}
\end{figure}

The overall average cross-band interference strength expressed in Equation~(\ref{eq:pcbi})
is determined by the following parameters: the interference link's power ($P_1$), 
the frequency separation to the interferer's $k$th subcarrier ($f-k$), 
the number of subcarriers occupied by the interference link ($L=|\Omega_1|$),
the overhead due to cyclic prefix ($\rho$) and the number of FFT/IFFT points  ($N$).
It is obvious that the cross-band interference increases linearly with the interference link's 
power $P_1$, and decreases quickly as the  frequency separation is enlarged.
But the impacts of $L$, $\rho$ and $N$ are implicit.
Next we seek to understand their impacts through numerical calculations,
and the results are shown in Figure~\ref{fig:comp_3para}.
We see  that the average cross-band interference strength slightly increases
as $L$ gets larger, and experiences a $3$dB reduction at the integer
frequency locations when $\rho$ changes from $0$ to $0.5$,
and it is almost insensitive to $N$ (as $N$ can be approximately cancelled).
Therefore, based on the above analysis we conclude that:
\begin{itemize}
\item Asynchronous transmissions create cross-band interference to each other
due to temporal mismatch and/or inter-link frequency offset;
\item Among those factors that have impacts on the average cross-band interference strength,
 the interference link's power and the frequency distance to the interference are dominating.
\end{itemize}

\section{Impact of Cross-Band Interference}
\label{sec:impact}
An OFDM packet consists of a  synchronization-targeted preamble and a number of data symbols.
In this section,
we focus on the impact of  the cross-band interference on packet detection/synchronization,
since it affects data reception  in frequency domain directly.

\subsection{Packet Detection and Synchronization}
In OFDM(A) systems, a receiver performs synchronization by detecting the packet preamble and
compensates the intra-link frequency offset~\cite{Robust97,Preamble00,Preamble01}. 
In most existing systems including 802.11a~\cite{802.11} and
WiMAX~\cite{802.16},  preambles are time-domain repeated symbols.
 To detect a preamble, 
the receiver uses a delay-correlation structure
to produce a peak when the preamble arrives.
The location of the peak marks the beginning of data frame,
and the phase
information extracted from the preamble  can be used to estimate the intra-link fractional frequency offset.
The integer frequency offset is usually compensated via another
pseudo-random noise (PN) sequence in frequency domain afterwards.

Two types of interference from the concurrent transmissions should be considered
with respect to packet synchronization. The first and stronger one is
the inter-channel interference caused by coexisting users operating on different frequency bands.
This type of interference can be effectively mitigated by adding band-pass filter, 
since the spectrum pattern is known to the receiver.
The second one is the cross-band interference
which can not be filtered out because it is the power leaked into the 
desired signal's main band. To examine the impact of cross-band interference
on synchronization performance, 
we assume that the inter-channel interference is fully 
filtered out. As to the evaluation metric,
we only examine the intra-link frequency offset error,
as the timing synchronization precision can be relaxed by adding long cyclic prefix.

\subsection{Impact on Packet Synchronization} 
We use the standard deviation of the intra-link frequency offset error
$\bigtriangleup f  $ (normalized to subcarrier spacing) and the  induced ICI to characterize the synchronization performance. 

We quote the preamble analysis in \cite{Robust97}. At high SNRs  without considering the interference, 
the standard deviation of $\bigtriangleup f  $ can be estimated as:
\begin{equation}
	{\mathrm {std}}[\bigtriangleup f  ]= \frac{\sqrt{2}}{\pi \sqrt{M\cdot \text{SNR}}}
\label{eq:fostd}
\end{equation}
where $M$ is the number of subcarriers occupied by the signal link,
$\text{SNR}$ is the average signal to noise power ratio (SNR) over all used subcarriers. 
To simplify our analysis, we assume the interference signal is Gaussian distributed
and independent of the desired signal and noise, so we
make a simple approximation by substituting
$\text{SNR}$ with $\text{SINR}$--the signal to interference-plus-noise
power ratio in Equation~(\ref{eq:fostd}),  \textit{i.e.},

\begin{equation}
\bigtriangleup f_{std}={\mathrm {std}}[\bigtriangleup f]\approx\frac{\sqrt{2}}{\pi \sqrt{M\cdot \text{SINR}}}.
\end{equation}
We as well assume ${\rm SINR}$ is high,
because it is measured over all used subcarriers
rather than at specific subcarriers.
Thus the average SINR is much lowered as compared with that
on the edge subcarriers near to the interferer. 
The interference power $P_I$ can be estimated in frequency domain:
\begin{equation}
	P_{I}=\frac{1}{M}\sum_{k=1}^{M} P_{CBI}(\epsilon+k)
\label{eq:PI}
\end{equation}
where $\epsilon$ is the forementioned inter-link frequency offset.

We next derive the  average signal power  of link 2. 
 Let $\Omega_2$ represent
the set of subcarriers occupied by 
link 2. Similarly to the analysis in Section~\ref{sec:oob}, we can compute link 2's power spectrum by:
\begin{equation}
P_{2\rightarrow 2}(f)=|H_{2\rightarrow 2}(f)|^2 \cdot |S_2(f) |^2
\label{eq:P22}
\end{equation}
where $H_{2\rightarrow  2} (f)$ is the frequency-domain channel response  of link 2,
$S_2(f)$ is the desired signal's spectrum:
\begin{equation}
	S_2(f)\!=\!\sum_{k\in\Omega_2}s_2(k)\frac{\sin[\pi(f\!-\!k)]}{N\sin[\frac{1}{N}\pi(f-k)]}
	e^{-i\pi(f\!-\!k)\frac{N\!-\!1}{N}}
\end{equation}
where $s_2(k)$ is the frequency domain signal sent by link 2's transmitter.
Assuming $E\big[|H_{2\rightarrow 2}(f)|^2\big]=1$
and $E\big[|s_2(k)|^2\big]=P_2$ for all $k\in\Omega_2$,
we can derive the average signal strength of link 2 
 by taking the average of Equation~(\ref{eq:P22}):
\begin{eqnarray}
\label{eq:link1}
	E\big[P_{2\rightarrow 2}(f)\big]
	=P_2\sum_{k\in\Omega_2}\frac{\sin^2[\pi(f-k)]}{N^2\sin^2[\frac{1}{N}\pi(f-k)]} .
\end{eqnarray}

The residual frequency synchronization error  $\bigtriangleup f  $ will lead to 
additional  ICI due to the destroyed subcarrier orthogonality.
For a given $f$ investigated at link 2's receiver, it deviates from the
right subcarrier index $l$ by $\bigtriangleup f  $, where $l\in\Omega_2$.
This will result in useful signal's power loss and unwanted ICI.
In this case, Equation~(\ref{eq:link1}) can be decomposed into two components:
\begin{equation}
\left\{
	\begin{aligned}
	&E\big[P_{2\rightarrow 2}(f)\big] = P_{SIG}(f) + P_{ICI}(f)\\
	&P_{SIG}(f)\!\!=\!\!\frac{P_2\sin^2[\pi(f-l)]}{N^2\sin^2[\frac{1}{N}\pi(f\!-\!l)]}\!\!
	=\!\!\frac{P_2\sin^2(\pi\bigtriangleup f  )}{N^2\sin^2(\frac{1}{N}\pi\bigtriangleup f  )}\\
	&P_{ICI}(f)=P_2\sum_{k\in\Omega_2, k\neq l}\frac{\sin^2[\pi(f-k)]}{N^2\sin^2[\frac{1}{N}\pi(f-k)]}
    \end{aligned}
	\right.
\label{eq:Psig_Pici}
\end{equation}
where $P_{SIG}(f)$ and $P_{ICI}(f)$ represent signal power and ICI power respectively.
We see that $P_{SIG}(f)$ is independent of $f$, and  $P_{ICI}$
is actually  insensitive to $f$ as well.
The detailed analysis of ICI can be found in \cite{IISC}.

\subsection{Comparing ICI with Cross-Band Interference}
\label{sec:ber}
The received data symbols
now experience two types of interference: the cross-band interference 
and the extra ICI due to $\bigtriangleup f  $. 
The average carrier to interference power ratio (CIR) determines
the performance of data retrieval  which is defined as:
\begin{equation}
	\text{CIR}(f)=\frac{P_{SIG}(f)}{P_{ICI}(f) + P_{CBI}(f)}
\label{eq:cir}
\end{equation}
where $P_{SIG}(f)$, $P_{ICI}(f)$ are defined in Equation~(\ref{eq:Psig_Pici})
and $P_{CBI}(f)$ is defined in Equation~(\ref{eq:pcbi}).
With Equation~(\ref{eq:cir}), we are able to compute
the bit error rate at specific subcarriers under various flat fading or non-fading channels  
using conventional analytical approaches~\cite{bk:digital_com,bk:fading}.
Difference from the cross-band interference,
ICI produces almost the same interference power to all relevant subcarriers,
because $\bigtriangleup f  $ applies to each subcarrier equivalently.
Note that CIR is dependent on the power heterogeneity $p_r$ which 
denotes the difference between $P_1$ and $P_2$.

ICI is much weaker than the worst case of the
cross-band interference, due to the robustness of the synchronization scheme.
 We use an example to show this property.
Let
link 1 (the interference link) and link 2 (the signal link) occupy the subcarriers 
\#$-7\to$\#$0$ and \#$1\to$\#$8$ out of 64 subcarriers respectively,
without inter-link frequency offset nor
power heterogeneity {(\em i.e.}, $\epsilon = 0$, $P_1=P_2$).
Also additive white Gaussian noise  is added with $P_n=-40$dB per subcarrier.
The analytical results are derived as follows.
The normalized $P_I$ can be estimated through Equation~(\ref{eq:PI}):
$P_{I}=\frac{1}{8}\sum_{k=1}^{8} P_{CBI}(k)=-15.1$dB.
Using this result we are able to calculate that ${\rm SINR}=1/(P_I+P_n)=15.1$dB 
and $\bigtriangleup f  _{std}\approx 0.028$ (normalized to subcarrier spacing).
Even using the adverse case of substituting $2\bigtriangleup f  _{std}$ for  $\bigtriangleup f $ in Equation~(\ref{eq:Psig_Pici}),
we still achieves tolerable signal loss and ICI: $P_{SIG}=-0.1$dB and $P_{ICI}\in(-23.7,-20.9)$dB. 
While the average cross-band interference strength at \#1 (strongest) and \#8 (weakest) are 
$P_{CBI}(1)=-9.1 {\rm dB}$ and $P_{CBI}(8)=-22.1 {\rm dB}$ respectively.
From this example we see that 
\begin{equation}
\left\{
\begin{aligned}
	&{\rm max}\{P_{CBI}(f)\} \gg P_{ICI}\\
	&{\rm min}\{P_{CBI}(f)\}\approx P_{ICI}
\end{aligned}
\right.
\end{equation}
So at the most concerned subcarriers that experience stronger cross-band interference interference, 
the synchronization error induced ICI can be ignored.

The negligibility of the impact of synchronization errors simplifies our analysis by  assuming perfect synchronization.
However, when multiple non-contiguous OFDM (NC-OFDM) spectrum blocks  are used by one user~\cite{ncofdm1_05,ncofdm2_05},
the cross-band interference from more neighbors would become stronger. In this case,
the impact of the cross-band interference on synchronization may not be negligible,
which we will study in our future work.

\section{Tackling Cross-Band Interference}
\label{sec:csc}

In this section, we introduce three mechanisms to
mitigate cross-band interference.
By adding frequency
redundancy, these mechanisms seek to reduce the
interference each subcarrier leaks to its neighbors, and/or to insert robustness at each receiver 
against such interference. The first two mechanisms are from prior works and the
last mechanism is our new proposal.

\subsection{Frequency Guardband (FGB)}
Inserting frequency guardband between transmissions is the most
straightforward way to mitigate cross-band interference~\cite{Window}. As shown in
Figure~\ref{fig:comp_sidelobe}, the interference strength decreases with the
frequency separation. For example, adding one subcarrier as guardband
between link 1 and 2 will reduce the interference power from
$-9.1$dB to $-13.5$dB. Such $4.4$dB difference could significantly improve transmission
quality. When a frequency guardband of $f_{gb}$ subcarriers is used between two
transmissions, the average cross-band interference strength can be derived from the original
analysis (Equation~(\ref{eq:pcbi})) by shifting $P_{CBI}$ by $f_{gb}$
subcarriers (become into $P_{CBI}(f+f_{gb})$). Because guardband is not used for transmissions, this approach
leads to extra overhead. 

The choice of guardband size depends heavily on 
power heterogeneity $p_r$ and the minimal CIR requirement ${\rm CIR}_{\rm min}$.
In table~\ref{tab:fgb}, we summarize the needed minimal guardband size (normalized to subcarrier spacing),
assuming $T_{CP}=T/4$ and 8 subcarriers are used by the interference link.
From the results we see that larger guardband is needed as the power heterogeneity gets higher.
This observation is helpful in designing spectrum allocation strategies.
\begin{table}[!h]
  \centering
  \caption{Minimal guardband size in terms of number of subcarriers}\label{tab:fgb}
  \begin{tabular}{|c|c|c|c|c|}
    \hline
    \backslashbox{${\rm CIR}_{\rm min}$}{$p_r$ }  & 0dB & 3dB & 6dB & 9dB \\
    \hline
     5dB & 0 & 0 & 0.6 & 1.6  \\
    \hline
     10dB & 0.2 & 1.0 & 2.0 & 4.0  \\
    \hline
     15dB & 1.8 & 3.7 & 5.9 & 10.0  \\
    \hline
  \end{tabular}
\end{table}

\subsection{Intra-Symbol Cancellation (ISC)}
Another mechanism is to reduce interference strength by carefully
controlling the  signal pattern. Prior work has proposed a series of
approaches~\cite{PCC,CC} where a transmitter codes several
subcarriers within one symbol in such a way that their signal sidelobes cancel each other
and hence the aggregated sidelobe becomes much smaller.
We call this type of interference mitigation methods  {\em intra-symbol
cancellation} (ISC). For example, if a transmitter
codes its subcarrier $k$ and $k-1$ such that:
\begin{equation}
s_1(k)=-s_1(k-1),
\end{equation}
then at any time-synchronized receiver, the cross-band interference from $k$ and
$k-1$ is expected to cancel each other out. 
In fact, when $\tau= 0$,  according to Equation~(\ref{eq:eq_A})
the sum power spectrum of the $k$th and $(k-1)$th subcarriers is

\begin{equation}
\begin{aligned}
|S_1(f)|_{k,k+1}^2&\!=\!\Big|\!\sum_{k,k+1}\!\!\big(s_1(k)\big)_0\!\frac{\sin[\pi(f\!\!-\! k)]\!\cdot \! e^{-i\pi(f\!-\!k)\frac{N\!-\!1}{N}}}{N\sin[\frac{1}{N}\pi(f\!\!-\!k)]}
 \Big|^2 \\
 \!\! &\approx \!P_1\!\Big[\frac{\sin[\pi(f\!\!-\! k)]}{\pi(f-k+1)(f-k)}\Big]^2
\end{aligned}
\end{equation}
In this case, the interference power decreases more quickly as compared with the non-mitigation case.

This approach, however, faces two key challenges. First, it leads to extra
overhead since two subcarriers now only carry the load of one
subcarrier.  Second and more importantly, ISC is sensitive to the
temporal mismatch. 
Specifically, in the presence of temporal mismatch, 
each subcarrier has a phase shift as we have shown in Section~\ref{sec:oob}.
The phase shift of $s_1(k)$ (in small mismatch) or $s_1^{I}(k)$, $s_{1}^{I\!I}(k)$ (in large mismatch)
depends on $k$ and $\tau$, thus the $k$th and $(k-1)$th subcarriers experience different phase shifts
and their sidelobes cannot effectively cancel each other. Mathematically deriving the interference strength 
is complex, so later we use simulation results to verify our intuitional analysis.

\subsection{Cross-Symbol Cancellation (CSC)}
To address the temporal mismatch, we propose to extend coding across
symbols,  referred to as the {\em cross-symbol
cancellation}~(CSC).  In contrast with ISC, CSC modifies every two
consecutive OFDM symbols on the same $k$th
subcarrier to compensate for the temporal mismatch.
Consider $s_1^{I}(k)$ and $s_{1}^{I\!I}(k)$ already defined above, after CSC, they become:
\begin{equation}
s_1^{I\!I}(k)=\big(s_1^{I}(k)\big)_{-T_{CP}}
\end{equation}
The factor  $(\cdot)_{-T_{CP}}$  compensates the phase
shift so that for both Case A and B, the actual interference signal
will have continuous phases, thus it can be seen as one completed OFDM symbol.
In doing so, no matter how large the temporal mismatch is,
the power spectrum of the $k$th subcarrier can be derived as
\begin{equation}
|P_{CBI}(f)|_{k}^2=P_1\frac{\sin^2[\pi(f- k)]}{N^2\sin^2[\frac{1}{N}\pi(f-k)]}
\end{equation}
 In this case, the aggregated signal sidelobe can be minimized
at integer frequency locations. Like ISC, CSC must trade off the
bandwidth overhead for interference mitigation. For example, coding a
small set of subcarriers on the frequency edge is more bandwidth-efficient than
coding all the subcarriers together, but is less effective in tackling the interference.

There are two steps to follow in the CSC-based system implementation.
1) At the transmitter, all links perform CSC coding on the edging $M$ subcarriers across every two successive OFDM symbols,
where $M$ is chosen according to the system requirements;
2) At the receiver, data symbols are jointly decoded through every two precoded symbols,
at least one of which will experience  the interferer's symbols with continuous phases of the $M$ subcarriers.
Choosing the less-interfered symbol from the two CSC-coded symbols 
can be assisted with channel estimation or other approaches, but we 
omit the details due to page limit.

Using the analytical results in the previous section, we verify that CSC
is insensitive to the temporal mismatch. More importantly,  the cross-band interference
strength $P_{CBI}(f)=0$ when $f (f\notin \Omega_1)$ is an integer of the
subcarriers with full-bits coding. That is, if the frequency offset is zero,
the desired signal is sampled at integer $f$
and does not experience any cross-band interference. On the other hand, the
performance of CSC is sensitive to the inter-link frequency offset. When the offset is
0.5 subcarrier, CSC suffers from almost the same interference as the original system.
Therefore, the proposed CSC is suitable for the frequency synchronized network.

\section{Evaluation}
\label{sec:eval}
In this section, we perform Matlab simulations to verify the accuracy of our
analytical models and further compare the effectiveness of the three
interference mitigation mechanisms.

\subsection{Experiment Setup}
By default, we simulate two links (link 1 and 2), each occupying 8 frequency
subcarriers out of a total of 64 subcarriers.  The links operate
asynchronously and start their transmissions randomly. Treating link 1 as the
interference link, we measure the impact of cross-band interference on link 2's
transmissions. We use four metrics: the interference strength at each
subcarrier, the BER at each subcarrier, the impact on packet synchronization
 and the overall effective link
throughput.  Similarly we also examine link 2's performance with and without
any interference mitigation mechanisms.

Table~\ref{tab:sim_para} summarizes the default simulation parameters.
We choose $T_{CP}=T/4$ which is the standard configuration in 802.11
systems. We also consider three practical artifacts: power heterogeneity,
channel fading and inter-link frequency offset.
To examine the impact of power heterogeneity $p_r$, we configure the interferer
strength to be $p_r$dB stronger than that of the desired signal at their intended
subcarriers, \textit{i.e.} $p_r=\text{log}(P_1/P_2)$.  We then vary $p_r$ to examine the
performance when the interferer is stronger or weaker than the target
signal.  To examine the impact of channel fading, we consider a Rician fading
environment and vary the $K$-factor to
control the weight of the line-of-sight and the Rayleigh fading components. 
When $K\rightarrow 0$, Rician fading becomes  Rayleigh fading, 
which has only scattered signal components. And when
$K\rightarrow +\infty$, it can be seen as non-fading.
For simplicity, we assume $H_{1\rightarrow 2}(f)=H_{2\rightarrow 2}(f)$.

\begin{table}[!t]
  \centering
  \caption{{ Default simulation parameters}}\label{tab:sim_para}
  \begin{tabular}{|l|l|}
    \hline
    {Parameter} & {Value} \\ \hline
    $N$ & 64 \\ \hline
    $N_{CP}$& 16 \\ \hline
    Modulation & QPSK \\ \hline
    Subcarrier bandwidth & 12.5 $k$Hz \\ \hline
    Desired signal's bandwidth & 8 subcarriers \\ \hline
    Interferer's bandwidth & 8 subcarriers \\ \hline
    Packet size & 32 OFDM symbols\\ \hline
    Guardband size& 0 subcarrier\\ \hline
  \end{tabular}
\end{table}

\begin{table}[!t]
  \small
  \centering
  \caption{Average interference strength under non-fading channel and Rayleigh fading channel}
  \label{tab:sim_err}
  \begin{tabular}{|c|c|c|c|c|c|c|c|}
   	\hline
    {$f$} & \multicolumn{3}{|c|}{$ P_{CBI}(f)$ {(dB)}}  & {$f$} &\multicolumn{3}{|c|}{$P_{CBI}(f)$ {(dB)}}   \\
    \cline{2-4} \cline{6-8}  & {NonFd} & {Ryl.} & {Theo.}  &  & {NonFd} & {Ryl.} & {Theo.} \\    \hline
    1 & $-9.0$  & $-9.2$  & ${\textbf{-9.1}}$  & 5 & $-19.2$ & $-19.0$ & $\textbf{-19.2}$\\ \hline
    2 & $-13.5$ & $-13.5$ & $\textbf{-13.5}$ & 6 & $-20.3$ & $-20.1$ & $\textbf{-20.3}$ \\ \hline
    3 & $-16.0$ & $-16.2$ & $\textbf{-16.1}$ & 7 & $-21.2$ & $-21.4$ & $\textbf{-21.3}$ \\ \hline
    4 & $-17.8$ & $-17.7$ & $\textbf{-17.8}$ & 8 & $-22.1$ & $-22.1$ & $\textbf{-22.1}$ \\
    \hline
  \end{tabular}
\end{table}

\subsection{Evaluating Cross-band Interference}
We start from examining the strength of cross-band interference and its impact
in terms of the packet synchronization and BER performance.

\para{Interference Strength Measurement.} By generating the 
link 1's signal and performing DTFT on the received signal at link 2, we
measure  the average interference power over 10,000 runs. This measurement
allows us to verify the accuracy of our analytical model by comparing it with
the experimental results. Two typical channels--non-fading channel and Rayleigh fading 
fading channel are used in our simulations. 
In Table~\ref{tab:sim_err} we list the average
signal strength derived from the simulation and our theoretical analysis,
where the two results differ by at most $0.2$dB. 
We also examined the maximum difference between the two when link 1 uses
different number of subcarriers, {\em i.e.\/} varying $L=|\Omega_1|$,
and the results match our analysis too.

\begin{figure}[!t]
	\centering
	\resizebox{0.48\textwidth}{!}{\includegraphics{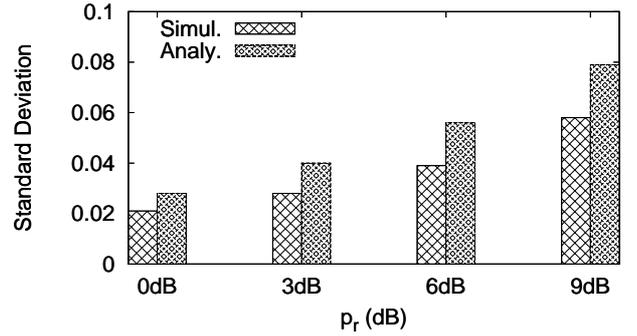}}
	\caption{Simulation results on intra-link frequency offset error.}
	\label{fig:sync_err}
\end{figure}

\para{Impact on Packet Synchronization.}
We evaluate the interference's impact
on packet synchronization by measuring the statistics of 
$\bigtriangleup f$--the intra-link frequency offset error.
Using a multi-band filter generated by  windowing method (\textit{e.g.}, Hamming),
link 1's main-band interference is filtered out and 
only  cross-band interference leaks into link 2's receiver.
Assuming  the actual intra-link frequency offset is randomly distributed in $[-0.5,0.5]$,
we run frequency offset estimation and measure the
standard deviation of $\bigtriangleup f$ over 10,000 simulations. 
Figure~\ref{fig:sync_err}
shows the measurement results, from which
we see that the simulation results have smaller
standard deviations than the analytical results in the presence of power heterogeneity.
One reason for that is the non-Gaussian distributed 
cross-band interference has weaker impact than the assumed Gaussian distributed one.
Assisted by the analysis in Section \ref{sec:ber}, it is reasonable to neglect the impact of the cross-band interference
on packet synchronization when evaluating system performance.

\begin{figure}[!t]
    \centering
		\subfigure[Impact of power heterogeneity]{\includegraphics[width=0.48\textwidth] 				  	
		{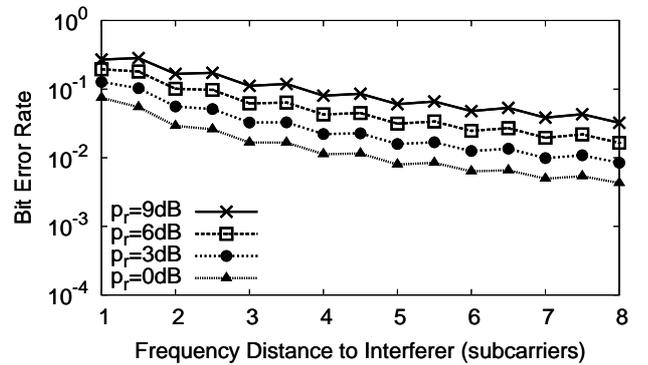}}
		\vfill
		\subfigure[Impact of channel fading properties]{\includegraphics[width=0.48\textwidth]
		{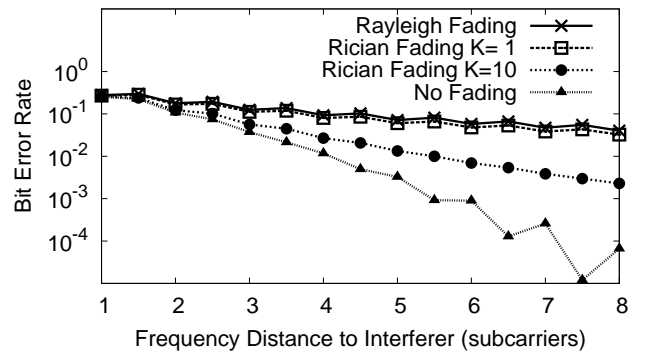}}
	\caption{Simulation results on BER performance measurement. (a) Impact of power heterogeneity in Rayleigh fading channel; (b) Impact of channel fading when $p_r=9$dB.}
    \label{fig:ber_fading}
\end{figure}

\para{BER Performance Measurement.}  Next, we assume no intra-link synchronization error 
and focus on examining the BER caused by the cross-band interference. 
We consider the power heterogeneity $p_r$ and Rician fading with $K$-factor
individually. 
We vary $p_r$ from $0$dB to $9$dB in Rayleigh fading channel ($K=0$),
and consider four channel fading
situations: $K=\infty$ (non-fading), 
$K=10$, $K=1$, and $K=0$  (Rayleigh fading) under large power heterogeneity when $p_r=9$dB. 
Figure~\ref{fig:ber_fading}  shows the simulation results.
We see that the BER at
each subcarrier increases gracefully as $p_r$ increases, and
increases dramatically as $K$ decreases (from non-fading to Rayleigh fading).

\begin{figure}[!t]
	\centering
	{\resizebox{0.48\textwidth}{!}{\includegraphics{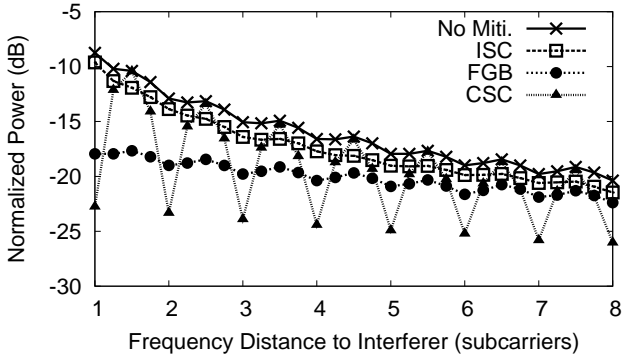}}}
	\caption{Simulation results on comparison of three mitigating approaches with 25\% overhead for FGB, CSC and ISC.}
	\label{fig:comp_interf}
\end{figure}

\begin{figure}[!t]
    \centering
	{\resizebox{0.48\textwidth}{!}{\includegraphics{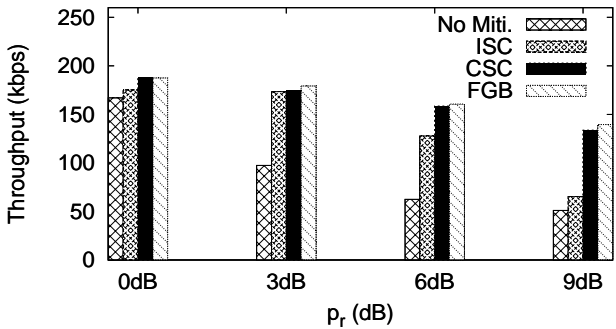}}}
	\caption{Simulation results on link throughput measurement with power heterogeneity. 
		$\epsilon$ is uniformly distributed in $[-0.1,0.1]$ subcarrier.}
	\label{fig:comp_thuput}
\end{figure}

\begin{figure}[!t]
	\centering
    {\resizebox{0.48\textwidth}{!}{\includegraphics{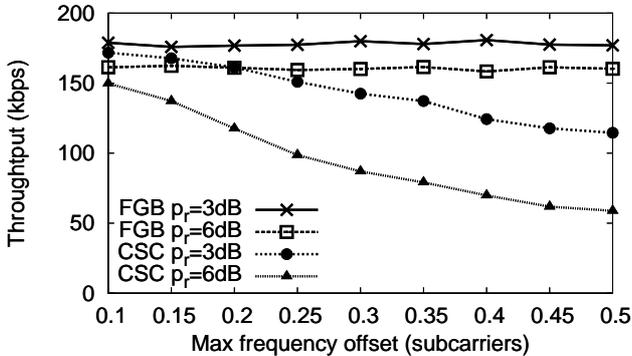}}}
    \caption{Sensitivity to inter-link frequency offset using CSC and FGB. The frequency
          offset is randomly distributed between $[-\epsilon_{max},\epsilon_{max}]$.}
    \label{fig:foff_impact}
\end{figure}

\subsection{Comparing Interference Mitigating Mechanisms}
Next, we examine the effectiveness of various interference mitigation
mechanisms, using both interference strength and link throughput
measurements.

\para{Interference Strength Measurement.}
We first compare the three mitigating methods--FGB, ISC and CSC with the
original system in terms of interference strength.
Figure~\ref{fig:comp_interf} shows the results when these mitigation methods
spend 25\%  overhead to combat the interference. As expected, ISC has
little improvement over the original system due to its sensitivity 
to the random temporal mismatch; CSC reduces the interference strength dramatically only at
integer frequency separations, showing its 
sensitivity to the inter-link frequency offset; while FGB reduces the interference
strength significantly, especially at subcarriers that are close to the
interferer.

\para{Link Throughput Measurement.}
We use link throughput measurements to examine the impact of cross-band
interference and the mitigation overhead. We
turn both link 1 and 2 on, randomly generate 10,000 OFDM packets, each
carrying 32 symbols, and measure link 2's effective throughput.  We
assume there is no intra-link synchronization error, but link
1 and 2  have a inter-link frequency offset.
We also use AWGN channel with $-40$dB additive white Gaussian noise.

Because the link throughput depends on the coding overhead in ISC/CSC
and the amount of frequency guardband in FGB, we configure the experiments to
perform a fair comparison. Assuming a total of 16 subcarriers, we choose the
best coding format for ISC/CSC to maximize link 2's effective throughput, using
the same configuration for link 1 and 2.
For FGB, we place $X$ null subcarriers between link 1 and 2, and
split the rest $16-X$ subcarriers between them. We choose the optimal $X$
that maximizes the effective throughput.

Figure~\ref{fig:comp_thuput} shows the throughput results where $p_r$ ranges
from $0$dB to $9$dB and  $\epsilon$ is randomly distributed in $[-0.1,0.1]$
subcarrier.  Both CSC and FGB consistently outperform the other two.
When the interferer is $9$dB stronger than the desired
signal, FGB and CSC provide roughly 100\% throughput improvement.  FGB
outperforms CSC slightly because CSC cannot nullify the
interference when $\epsilon\neq 0$. The difference between FGB and CSC is
small because  $\epsilon$ is relatively small in our experiments.

\para{Impact of Inter-link Frequency Offset.} To examine the impact of inter-link frequency offset,
on compare CSC and FGB when the inter-link frequency offset is uniformly distributed
in $[-\epsilon_{max}, \epsilon_{max}]$. Results from Figure~\ref{fig:foff_impact} show that the link
 throughput of CSC degrades gracefully with $\epsilon_{max}$, especially when the
desired signal is weaker. This
can be explained by Figure~\ref{fig:comp_interf} where the CSC's cross-band
interference increases with $\epsilon$. When $\epsilon_{max}=0.5$ subcarrier,
the interference becomes similar to the no-mitigation case.  On the other hand FGB is almost insensitive to $\epsilon_{max}$.

\section{Conclusions and Future Work}
\label{sec:conclusion}
In this paper, we consider the problem of distributed spectrum sharing using
OFDMA. 
We show that artifacts from asynchronous transmissions destroy subcarrier orthogonality, 
creating cross-band interference among transmissions. 
We develop an analytical framework to
quantify the strength of cross-band interference and evaluate 
its impacts on preamble detection as well as
data reception.  We show that the cross-band interference is present and can
produce large performance degradations. 
We then build and compare three methods to mitigate the interference. 
Experimental and
analytical results show that adding frequency guardband is the most efficient
solution in the presence of temporal mismatch and frequency offset.
While choosing the guardband size depends heavily on 
the power heterogeneity between spectral adjacent links.

We are considering extending
our work in the following directions. 
First, 
we are currently studying the impact of user heterogeneity, 
such as power heterogeneity, symbol length heterogeneity {\em etc.}, on the cross-band interference.
Second, our analysis 
considers flat fading channel in this paper, which can be
extended to frequency selective fading channel. Last, using insights of 
our analysis, it is worthwhile investigating how to integrate the 
interference mitigation methods into practical dynamic spectrum access protocols.

\appendix
\section{}
\label{sec:append}
\allowdisplaybreaks
In the following derivations, we denote $t_1(n)$, $t_1^{I}(n)$ and $t_1^{I\!I}(n)$ 
as the time-domain sampling points of link 1's signal seen from link 2's receiver 
(see Figure~\ref{fig:sampling}). $t_1(n)$ refers to the samples in small
mismatch case, and
$t_1^{I}(n)$ and $t_1^{I\!I}(n)$ refer to the samples of Symbol I and Symbol II in large mismatch case, respectively.

\subsection{Derivation of $S_1(f)$ in Eq.~$(\ref{eq:eq_A})$.}

\begin{eqnarray}
{S}_1(f) &=& {DTFT}[t_1(n)] \nonumber \\
&=& \sum_{n=0}^{N-1}t_1(n)e^{-i2\pi f\frac{n}{N}} \nonumber \\
&=& \sum_{n=0}^{N-1}{IDFT \big[\big(s_1(k)\big)_{\tau}\big]} \cdot e^{-i2\pi f\frac{n}{N}}\nonumber\\
&=& \sum_{n=0}^{N-1} \Big\{ \frac{1}{N}\sum_{k\in\Omega_1}\big(s_1(k)\big)_{\tau} \cdot e^{i2\pi k \frac{n}{N}} \Big\} e^{-i2\pi f\frac{n}{N}}\nonumber\\
&=& \sum_{k\in\Omega_1}\big(s_1(k)\big)_{\tau} \Big\{ \frac{1}{N}\sum_{n=0}^{N-1} \cdot e^{-i2\pi (f-k)\frac{n}{N}}\Big\}\nonumber\\
&=& \sum_{k\in\Omega_1}\big(s_1(k)\big)_{\tau}\frac{\sin[\pi(f-k)]}{N\sin[\frac{1}{N}\pi(f-k)]} \nonumber\\
&&  \qquad \cdot e^{-i\pi(f-k)\frac{N-1}{N}} \nonumber
\end{eqnarray}
Note that $\big(s_1(k)\big)_{\tau}$ reflects the relative cyclic shift of $t_1(n)$ by $\tau$ compared with link 2's signal.

\subsection{Derivation of $S_1^{I}(f)$ and $S_1^{I\!I}(f)$ in Eq.~$(\ref{eq:eq_B})$}

\begin{eqnarray}
{S}_{1}^{I}(f) &=& {DTFT}[t_1^{I}(n)] \nonumber \\
&=& \sum_{n=0}^{M-1}t_1^{I}(n)e^{-i2\pi f\frac{n}{N}} \nonumber \\
&=& \sum_{n=0}^{M-1}{IDFT \big[\big(s_1^{I}(k)\big)_{\tau}\big]} \cdot e^{-i2\pi f\frac{n}{N}}\nonumber\\
&=& \sum_{n=0}^{M-1} \Big\{ \frac{1}{N}\sum_{k\in\Omega_1}\big(s_1^{I}(k)\big)_{\tau}e^{i2\pi k \frac{n}{N}} \Big\} \cdot e^{-i2\pi f\frac{n}{N}}\nonumber\\
&=& \sum_{k\in\Omega_1}\big(s_1^{I}(k)\big)_{\tau} \Big\{ \frac{1}{N}\sum_{n=0}^{M-1} \cdot e^{-i2\pi (f-k)\frac{n}{N}}\Big\}\nonumber\\
&=& \sum_{k\in \Omega_1} \big(s_1^{I}(k)\big)_{\tau} \frac{\sin[\frac{M}{N}\pi(f-k)]}{N\sin[\frac{1}{N}\pi(f-k)]}\nonumber\\
&&  \qquad \cdot e^{-i\pi(f-k)\frac{M-1}{N}} \nonumber
\end{eqnarray}

\begin{eqnarray}
{S}_{1}^{I\!I}(f) &=& {DTFT}[t_1^{I\!I}(n)] \nonumber \\
&=& \sum_{n=M}^{N-1}t_1^{I\!I}(n)e^{-i2\pi f\frac{n}{N}} \nonumber \\
&=& \sum_{n=m}^{N-1}{IDFT \big[\big(s_1^{I\!I}(k)\big)_{\tau}\big]} \cdot e^{-i2\pi f\frac{n}{N}}\nonumber\\
&=& \sum_{n=M}^{N-1} \Big\{ \frac{1}{N}\sum_{k\in\Omega_1}\big(s_1^{I\!I}(k)\big)_{\tau}e^{i2\pi k \frac{n}{N}} \Big\} \cdot e^{-i2\pi f\frac{n}{N}}\nonumber\\
&=& \sum_{k\in\Omega_1}\big(s_1^{I\!I}(k)\big)_{\tau} \Big\{ \frac{1}{N}\sum_{n=M}^{N-1} \cdot e^{-i2\pi \frac{(f-k)n}{N}}\Big\}\nonumber\\
&=& \sum_{k\in\Omega_1}\big(s_1^{I\!I}(k)\big)_{\tau} \frac{\sin[\frac{N-M}{N}\pi(f-k)]}{N\sin[\frac{1}{N}\pi(f-k)]}\nonumber\\
&&  \qquad\cdot e^{-i\pi(f-k)\frac{N+M-1}{N}} \nonumber
\end{eqnarray}
Similarly, $t_1^{I}(n), t_1^{I\!I}(n)$ refer to the sampled points in the time domain,
 and $s_1^{I}(k), s_1^{I\!I}(k)$ refer to the symbols in the frequency domain.

\bibliographystyle{elsarticle-num}  
\bibliography{wei}

\end{document}